\documentclass[twocolumn,showpacs,showkeys,preprintnumbers,amsmath,prd,amssymb,superscriptaddress,chaproman]{revtex4-1}

\usepackage{graphicx}
\usepackage{dcolumn}
\usepackage{bm}
\usepackage{multirow}
\usepackage{color}%
\usepackage{epsfig}%
\usepackage{hyperref}
\hypersetup{
colorlinks=true,
citecolor=blue,
linkcolor=blue,
urlcolor=blue
}
\usepackage[english]{babel}
\usepackage{amsmath,amssymb}
\usepackage{amsfonts}
\usepackage{stmaryrd}
\usepackage{mathrsfs}
\usepackage{xcolor}
\usepackage{color}

\newcommand{\Eq}{Eq.~}

\def\nuebar{{\rm \bar{\nu}_e}}
\def\nuebare{{\rm \bar{\nu}_{e}-e}}
\def\nue{{\rm \nu_e}}
\def\nuee{{\rm \nu_{e}-e}}
\def\s2tw{{\rm sin ^2 \theta_{W}}}

\def\elr{{\rm \varepsilon_{ee}^{eL,R}}}

\def\etlr{{\rm \varepsilon_{e\tau}^{eL,R}}}

\def\Lambdaup{\Lambda_{\cal U}}
\def\dsca{d_{\cal S}}
\def\dvec{d_{\cal V}}

\begin{document}

\title{Constraints on scalar-pseudoscalar and tensorial
nonstandard interactions and tensorial unparticle
couplings from neutrino-electron scattering}

\newcommand{\deu}{Department of Physics,
Dokuz Eyl\"{u}l University, Buca, \.{I}zmir TR35160, Turkey}
\newcommand{\as}{Institute of Physics, Academia Sinica, Taipei 115, Taiwan}
\newcommand{\bhu}{Department of Physics, Banaras Hindu University, Varanasi 221005, India}
\newcommand{\metu}{Department of Physics,
Middle East Technical University, Ankara TR06800, Turkey}
\newcommand{\corr}{muhammed.deniz@deu.edu.tr}

\author{ M.~Deniz }  \altaffiliation [Corresponding Author: ]{ \corr }
\affiliation{ \deu } \affiliation{ \as }
\author{ B.~Sevda }  \affiliation{ \deu } \affiliation{ \as }
\author{ S.~Kerman }  \affiliation{ \deu } \affiliation{ \as}
\author{ A.~Ajjaq }  \affiliation{ \deu }
\author{ L.~Singh }  \affiliation{ \as } \affiliation{ \bhu}
\author{ H.T.~Wong }  \affiliation{ \as }
\author{ M.~Zeyrek }  \affiliation{ \metu }

\date{\today}

\begin{abstract}

Neutrino-electron scattering is a purely leptonic fundamental interaction
and therefore provides an important channel to test the Standard Model,
especially at the low energy-momentum transfer regime.
We derived constraints on neutrino nonstardard interaction couplings
depending on model-independent approaches which are described by a four-Fermi
pointlike interaction and unparticle physics model with tensorial components.
Data on $\nuebare$ and $\nuee$ scattering from the TEXONO and LSND experiments,
respectively, are used. The upper limits and the allowed regions of scalar,
pseudoscalar, and tensorial nonstandard interaction couplings of neutrinos are 
derived at 90\% confidence level in both one-parameter and two-parameter 
analysis. New upper limits for tensorial unparticle physics coupling constants 
and mass parameters are also placed.

\end{abstract}

\medskip


\maketitle

\section{INTRODUCTION}

Neutrinos play crucial roles in particle physics and cosmology. 
Discovery of neutrino oscillations shows that neutrinos have finite mass. 
There are intense experimental efforts to study their properties and 
interactions with matter. Therefore, they are keystones for completeness 
of the Standard Model (SM)~\cite{king2015}.
Solar and atmospheric experimental data have confirmed that neutrinos
do oscillate, thereby they are massive and can be mixed.
Extra new interactions due to nonstandard properties
of neutrinos which are often called nonstandard interactions (NSIs)
of a neutrino have not been observed experimentally yet,
mainly due to poor experimental sensitivities.

Recent and upcoming neutrino experiments will provide more precise measurements
on intrinsic neutrino properties~\cite{jpanman-wjmarciano}, and therefore have
the potential to open a new window for the observation of NSI effects~\cite{ohlsson2013}.
Nonoscillation experiments that have measured a neutrino cross section with high
accuracy may provide profound information for neutrino interactions resulting in
direct measurements of NSIs. These interactions are important not only for
phenomenological~\cite{miranda2015} but also for the experimental points of view
since the measurements and found evidence can suggest new physics or favor
one of the existing new physics theories beyond the SM (BSM).

Neutrino-electron scattering provides quite convenient channel for testing
the SM of electroweak theory, especially in
low-energy regime since it is a pure leptonic process
~\cite{jerler125,mdeniz033004,sbilmis073011,jiunn011301,sbilmis033009}.
Advanced systems capable of making measurements at low energy and low
background are necessary to observe neutrino interactions with good
experimental precision. In principle, there are some advantages to
studying with reactor neutrinos ($\nuebar$): the reactors are excellent
sources for a low-energy electron-type antineutrino with high neutrino
flux up to around 10 MeV. Besides reactor on and off
comparison providing a model-independent way of background
subtraction, the reactor $\nuebar$ spectra are understood
and well known. Therefore, these advantages provide better
experimental sensitivities.

This paper is a follow-up of our earlier studies on
(i) nonuniversal or flavor-conserving (FC) and
flavor-changing or flavor-violating (FV) NSI of neutrino,
and (ii) vector and scalar unparticle physics (UP)~\cite{mdeniz033004}.
We report experimental constraints on scalar, pseudoscalar, and tensorial
NSIs and tensorial unparticle couplings via neutrino-electron
elastic scattering interaction channels.

\section{NEUTRINO-ELECTRON SCATTERING AND DATA}
\subsection{Standard Model}

$\nue(\nuebar)-e$ elastic scattering can occur via both
charged current and neutral current.
Therefore, their interference which is destructive also contributes
to the cross section. The SM differential
cross section of $\nu_e(\bar{\nu}_e) - e$ elastic
scattering can be expressed in terms of the chiral coupling
of $g_{L}$ and $g_{R}$ in the laboratory frame
as~\cite{bkayser87,mdeniz033004}
\begin{eqnarray}
\left[ \frac{d\sigma}{dT}(\nu_{e}e ) \right] _\text{SM}  &=&
\frac{2G_{F}^{2}m_{e}}{\pi } \left[(g_{L}+1)^{2}
+g_{R}^{2}\left(1- \frac{T}{E_{\nu }}\right) ^{2}\right. \nonumber  \\
&-& \left.g_{R}(g_{L}+1)\frac{m_{e}T} {E_{\nu}^{2}} ~ \right]~,
\label{eq::nue_gr_gl}
\end{eqnarray}
and
\begin{eqnarray}
\left[ \frac{d\sigma}{dT}(\bar{\nu}_{e}e ) \right] _\text{SM}  &=&
\frac{2G_{F}^{2}m_{e}}{\pi } \left[g_{R}^{2}
+ (g_{L}+1)^{2}\left(1- \frac{T}{E_{\nu }}\right) ^{2}\right. \nonumber  \\
&-& \left.g_{R}(g_{L}+1)\frac{m_{e}T} {E_{\nu}^{2}} ~ \right]~,
\label{eq::nubare_gr_gl}
\end{eqnarray}
where $G_F$ is the Fermi coupling constant, $T$ is the kinetic
energy of the recoil electron, $E_\nu$ is the incident
neutrino energy, and $g_L = -\frac{1}{2} + sin^2{\theta_W}$
and $g_R = sin^2{\theta_W}$ are the chiral coupling constants
in terms of weak mixing angle $sin^2{\theta_W}$.

\subsection{Input data}

Short-baseline neutrino experiments provide some advantages to study BSM.
Because of the minimizing oscillation effect, short-baseline reactor neutrino 
experiments, where pure $\bar{\nu}_e$ is produced, unlike the mixing of 
different eigenstates of neutrinos as in the case of Solar and atmospheric ones, 
can be used to probe BSM effectively. Reactors produce high $\bar{\nu}_e$ fluxes 
compared to other sources. The reactor-off period provides a model-independent 
means of background subtraction. The studies of reactor $\bar{\nu}_e-e$ 
interaction provide better sensitivities to the SM electroweak parameters 
$\sin^2{\theta_W}$ and $g_V, g_A$ at the same experimental accuracies as those 
from $\nu_e$ measurements~\cite{mdeniz072001}. The lower neutrino energy at the 
MeV range also favors applications where sensitivities can be enhanced at
low detector thresholds.

In this paper, the analysis is based on data from:
(i) the TEXONO experiment on antineutrino-electron interactions at
low energy  using three different detectors located at Kuo-Sheng
Reactor Neutrino Laboratory (KSNL) and (ii) the LSND experiment
on neutrino electron interactions at high energy using accelerator neutrinos.
The results from three independent data sets
from TEXONO of $\nuebare$ interaction
are compared with those from the data set of LSND $\nuee$ interaction.

KSNL is located at a distance of 28 m from
one of the cores of Nuclear Power Plant in Taiwan with 30 m 
water-equivalent overburden.
The 2.9 GW reactor cores are produced an average
$\bar{\nu}_e$ flux of $\phi(\bar{\nu}_e)\sim 6.4\times10^{12}\ 
\text{cm}^{-2} \text{s}^{-1}$ at the experimental site. 
Detectors has been placed in a shielding structure with 
50 ton of passive materials and surrounded by active 
anti-Compton detectors: Cs(Tl) or NaI(Tl) for the anti-Compton 
detector, and a cosmic-ray veto scintillator array.

\begin{description}

\item[TEXONO Experiment] Three experimental data sets taken with
different detectors are used as follows:

\begin{description}

\item[\bf CsI(Tl).] $- 29882/7369$ kg-days of reactor on/off data:
$\bar{\nu}_e-e^-$ electroweak interaction cross section,
$g_V$, $g_A$, weak mixing angle $sin^2{\theta_W}$ and
charge radius squared were measured with an effective mass
of 187 kg CsI(Tl) crystal scintillator array
at $3-8$ MeV. The rms energy resolutions are 5.8\%,
5.2\%, and 4.0\% at $^{137}$Cs, $^{40}$K, and $^{208}$Tl
$\gamma$ peaks, respectively~\cite{mdeniz072001}.

\item[\bf HP-Ge.] $- 570.7/127.8$ kg-days of reactor on/off data:
The limit on the neutrino magnetic moments at 90\% C.L. was derived 
and the constraints on the couplings of axion were placed using a 
high-purity germanium detector with a target mass of 1.06 kg and 
measurements at 12–64 keV~\cite{hbli-htwong}.
The rms energy resolution of HP-Ge is
880 keV at Gallium K-shell x-ray energy~\cite{soma2016}.

\item[\bf PC-Ge.] $- 124.2/70.3$ kg-days of Reactor on/off data:
New limits are set to neutrino millicharge and low-mass weakly
interacting massive particles with a fiducial mass of 500 g point 
contact germanium (PC-Ge) detector~\cite{jiunn011301} in the 
$0.3-12$ keV energy region. The rms energy resolution of PC-Ge is
87 keV at Gallium K-shell x-ray energy~\cite{soma2016}.

\end{description}

\item[\bf LSND Experiment] The Liquid Scintillator Detector at
the Los Alamos Neutron Science Center uses neutrinos produced
at the proton beam stop with $T$ of $~18-50$ MeV.
The cross section for the elastic scattering reaction
$\nu_e-e$ and weak mixing angle $sin^2{\theta_W}$
were measured. The energy resolution was determined
from the shape of the electron energy spectrum and
was found to be 6.6\% at the 52.8 MeV end
point~\cite{auerbach2001}.
\end{description}

\subsection{Analysis methods}

FC and FV NSIs, as well as scalar and vector UP were studied in our previous
studies by using the data sets of CsI(Tl) and HP-Ge~\cite{mdeniz033004}.
In this paper, we adopt the same analysis methods for the new interaction
channels. A new data set of PC-Ge of TEXONO is also included for the analysis
to cover the lower energy range as well.

The expected event rate of $R$ can be calculated as

\begin{equation}
R_{X} ~ = ~ \rho_e ~ \int_{T} \int_{E_{\nu }}
\left[\frac{d\sigma}{dT}\right]'_{X}  ~ \frac{d \phi ( \nuebar ) }{dE_{\nu}} ~
dE_{\nu} ~ dT ~ ~,
\label{eq::RX}
\end{equation}
where $\rho_e$ is the electron number density per kilogram of target mass
and $d\phi/dE_\nu$ is the neutrino spectrum. The measurable differential
cross section is denoted by $\left[d\sigma/dT\right]'$,
which corresponds to convoluting the detector energy resolution to the physics
differential cross section $\left[d\sigma/dT\right]$.  In practice,
for the BSM models and experimental data studied in this work, the variations
of $\left[d\sigma/dT\right]$ with energy are gradual, such that
the resolution smearing does not significantly alter the measured spectra
in the region of interest. The difference between $\left[d\sigma/dT\right]$
and $\left[d\sigma/dT\right]'$  is less than  0.1\%. Accordingly,
resolution effects can be neglected in this analysis.

$R_{expt}$ and $R_{X}$ correspond to observed and expected event rates,
respectively. $X$ represents different interaction channels such as
the SM, NSI, etc. $R_{expt}$ is expressed in the unit of 
$\mbox{kg}^{-1} \mbox{MeV}^{-1} \mbox{day}^{-1}$ and $\mbox{kg}^{-1} 
\mbox{keV}^{-1} \mbox{day}^{-1}$ for CsI(Tl) and Ge data sets, respectively.

The results on physics couplings from this analysis are
expressed either as "best-fit $\pm$ statistical $\pm$
systematic uncertainties" at 1-$\sigma$ level, or in terms of
limits at 90\% C.L.. The statistical uncertainties are derived
by a minimum $\chi^2$ method defined as

\begin{equation}
\chi^2 = \sum\limits_{i=1} {\left[\frac{R_{expt}(i)-R_{SM}(i)-R_{X}(i)}
{\Delta(i)}\right]}^2 ~,
\label{eq::chi}
\end{equation}
where $R_{expt}$, $R_{SM}$ and $R_{X}$ are the measured event rate, SM and $X$
(= NSI, UP, etc.) expected event rates, respectively, while $\Delta(i)$
is the  statistical uncertainty of the ith bin published by the experiments.
The 1-$\sigma$ statistical errors in the physics couplings correspond to
those values that produce $\chi^2 = \chi_{min}^2 + 1$. The published systematic 
uncertainties of the experiments contribute to shifts of the best-fit values 
in the physics couplings. The two contributions are added in quadrature to 
give rise to the combined uncertainties, from which the 90\% C.L. limits 
can be  derived using the prescription of Ref. ~\cite{Gary98}.

\begin{figure}[!ht]
  \begin{center}
  \includegraphics[width=8cm]{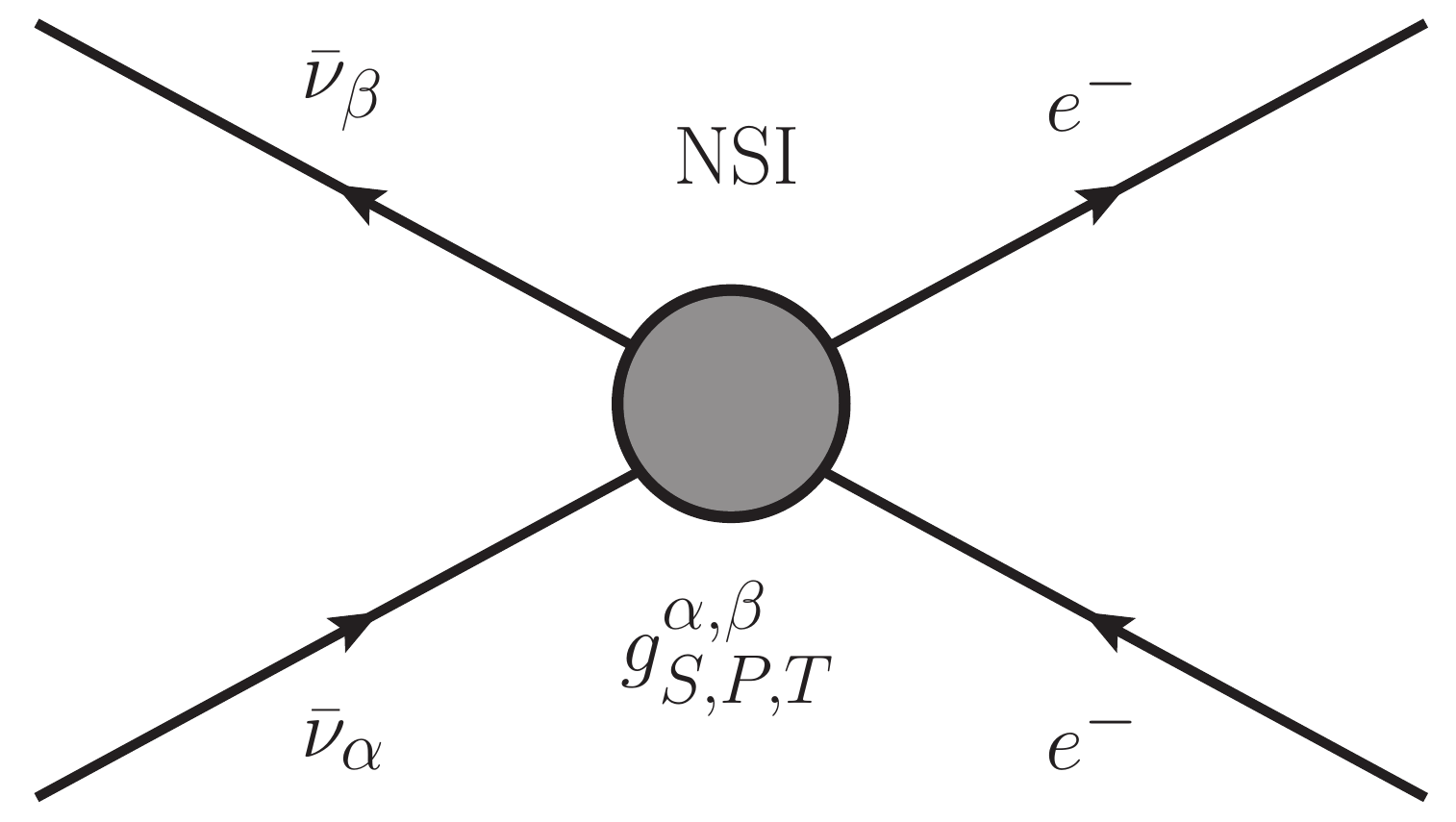}\\[4ex]
  \caption{NSIs of neutrinos, generically described as four-Fermi
  interaction with new couplings.}
  \label{fig::feydiag_nsi}
  \end{center}
\end{figure}
\begin{figure}[!ht]
  \begin{center}
  \includegraphics[width=8cm]{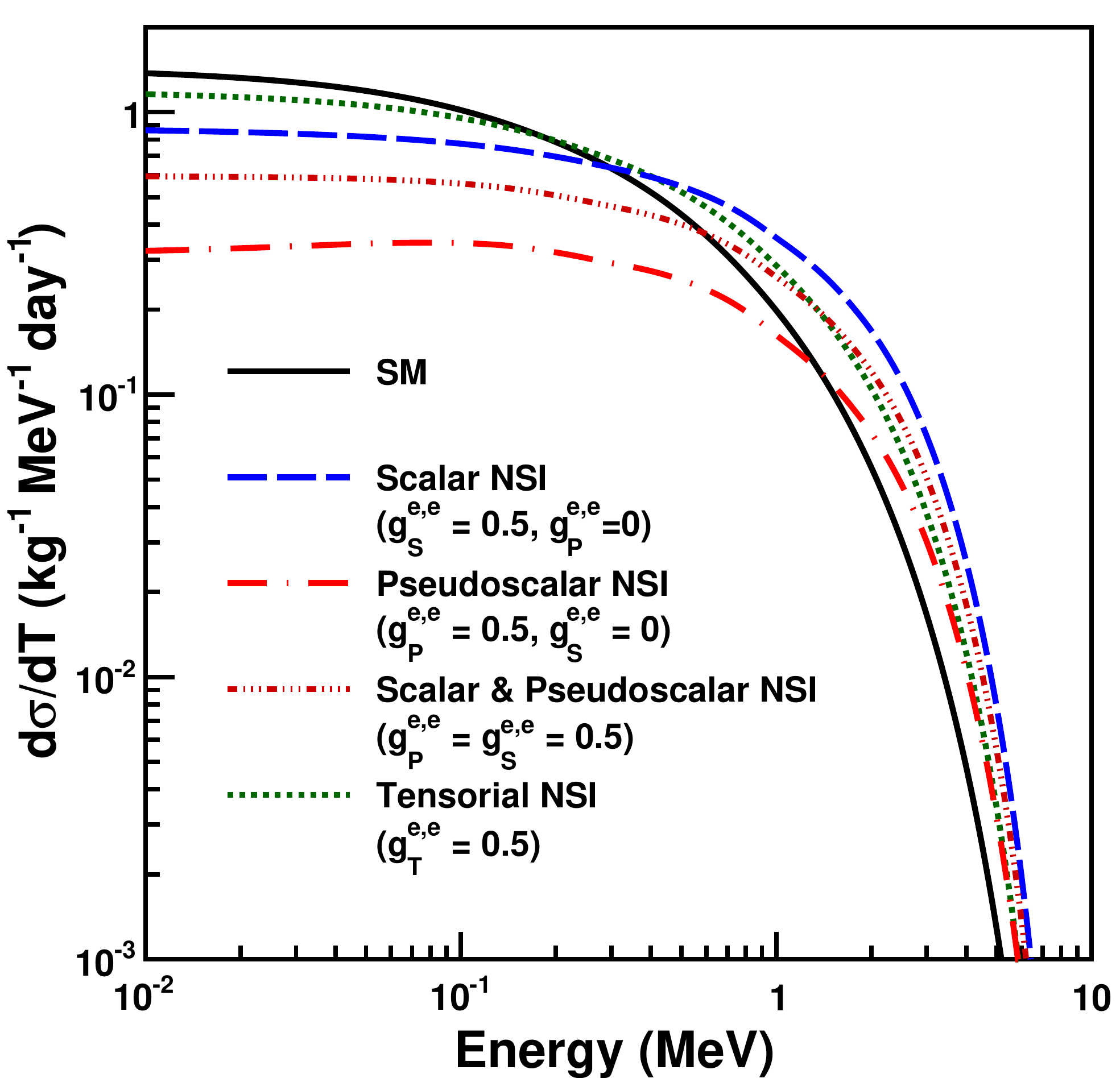} \qquad\qquad
  \caption{Differential cross section as a function of
  the recoil energy $T$ with typical reactor $\nuebar$ spectra
  for scalar, pseudoscalar and tensorial NSIs at some specific coupling 
  parameters using CsI(Tl) as a target.}
  \label{fig::diff_nsi}
  \end{center}
\end{figure}

\section{SCALAR, PSEUDOSCALAR and TENSORIAL NSI OF NEUTRINO}

Since neutrino-electron scattering is a pure leptonic process, it provides
 a very convenient channel to test the SM. NSI of neutrinos is first
considered as an alternative mechanism for neutrino oscillation. However,
NSI is now only allowed for lower pioneers effects to the neutrino oscillation
and can be used to improve the sensitivities of oscillation experiments.
In this paper, investigation of some of the BSM new physics scenarios
via $\nu_e(\bar{\nu}_e)-e$ elastic scattering is introduced
and the results will be given in subsequent sections.
In these new physics scenarios,
the NSI of neutrinos is considered as a model-independent interaction,
which is described as a four-Fermi pointlike or
so called zero-distance interaction.
Feynman diagram of different NSI for neutrino-electron
scattering is illustrated schematically in
Fig.~\ref{fig::feydiag_nsi}.

Both neutrino oscillation and non-oscillation experiments are sensitive to NSI
parameters and can give complementary results. Non-oscillation neutrino experiments
provide direct measurement of NSI while neutrino oscillation experiments are more
sensitive to propagation of NSI parameters due to matter effects~\cite{miranda2015}.
NSI can simply be considered as a modification of chiral coupling constants 
of $g_{L,R}$ with additional new physics terms,  in general. 

Phenomenological studies of FC and
FV NSIs of neutrinos have been extensively carried
out with a variety of interaction channels and neutrino sources
~\cite{nsiastrophys,nsiboundlsnd,nsiboundcombined,nsinuN,reactornubsm,cbig}.
Experimentally new bounds for FC coupling of
$\elr$ and FV coupling of $\etlr$
NSI parameters were derived and existing bounds were improved in our
earlier work by taking advantage of neglecting oscillation effects and
high neutrino flux~\cite{mdeniz033004}. On the other hand, other
NSIs of neutrinos are also possible and are of the
scalar, pseudoscalar, and spin-2 tensorial types.

\begin{table*} [!hbt]
\caption{Summary of best-fit results and corresponding limits
at \%90 C.L. for scalar, pseudoscalar, and tensorial NSI
measurements for one-parameter-at-a-time analysis for
$\nue -$ and $\nuebar - e$ scattering.}
\label{tab::gs_gp}
\begin{ruledtabular}
\begin{tabular}{cccccccc}
& \multicolumn{3}{c}{TEXONO} &
\multicolumn{2}{c}{LSND}\\
NSI & Measurement & \ & Bounds & Measurement & Bounds \\
Parameters & Best fit (1-$\sigma$) & $\chi^2$/dof & at 90\% C.L. 
& Best fit (1-$\sigma$) & at 90\% C.L.  \\ \hline \\

Scalar & $g_S^{e,e}=$ & 8.7/9 & $-0.317 <$ & $g_{S}^{e,e}=$ & $-0.880 <$ \\

$g_S^{e,e}\ (g_P^{e,e}=0)$ & $\left[3.27 \pm 6.39 \pm 3.10 \right]\times10^{-2}$
& \ & $g_{S}^{e,e} < 0.113$ & $0.27 \pm 0.59 \pm 0.26$ & $g_{S}^{e,e}<0.642$ \\ \\

Pseudoscalar & $g_P^{e,e}=$ & 8.7/9 & $-0.113 <$ & $g_P^{e,e}=$ & $-0.642<$ \\

$g_P^{e,e}\ (g_S^{e,e}=0)$ & $\left[-3.27 \pm 6.39 \pm 3.10 \right]\times10^{-2}$
& \ & $g_{P}^{e,e} < 0.317$ & $-0.27 \pm 0.59 \pm 0.26$ & $g_{P}^{e,e}<0.880$ \\ \\

$g_{S=P}^{e,e}\ (g_S^{e,e}=g_P^{e,e})$ & ${\left(g_{S=P}^{e,e}\right)}^2=$ & 8.7/9 &
$|g_{S=P}^{e,e}| < 0.100$ & ${\left(g_{S=P}^{e,e}\right)}^2=$ & $|g_{S=P}^{e,e}| < 0.375$ \\

\ & $\left[0.19 \pm 0.38 \pm 0.31 \right]\times10^{-2}$ & \ & \ &
$\left[3.47 \pm 4.78 \pm 4.36 \right]\times10^{-2}$ & \ \\ \\

Tensorial & ${\left(g_{T}^{e,e}\right)}^2=$ & 8.7/9 & $|g_{T}^{e,e}| < 0.238$ &
${\left(g_{T}^{e,e}\right)}^2=$ & $|g_{T}^{e,e}| < 0.401$ \\

$g_T^{e,e}$ & $\left[0.96 \pm 2.21 \pm 1.82 \right]\times10^{-2}$ & \ & \ &
$\left[3.96 \pm 5.47 \pm 4.97 \right]\times10^{-2}$ & \ \\ \\
\end{tabular}
\end{ruledtabular}
\end{table*}
\begin{figure*}[!ht]
\begin{center}
{\bf \hspace{0.1cm}(a)} {\bf \hspace{8.5cm}(b)} \\
  \includegraphics[width=8cm]{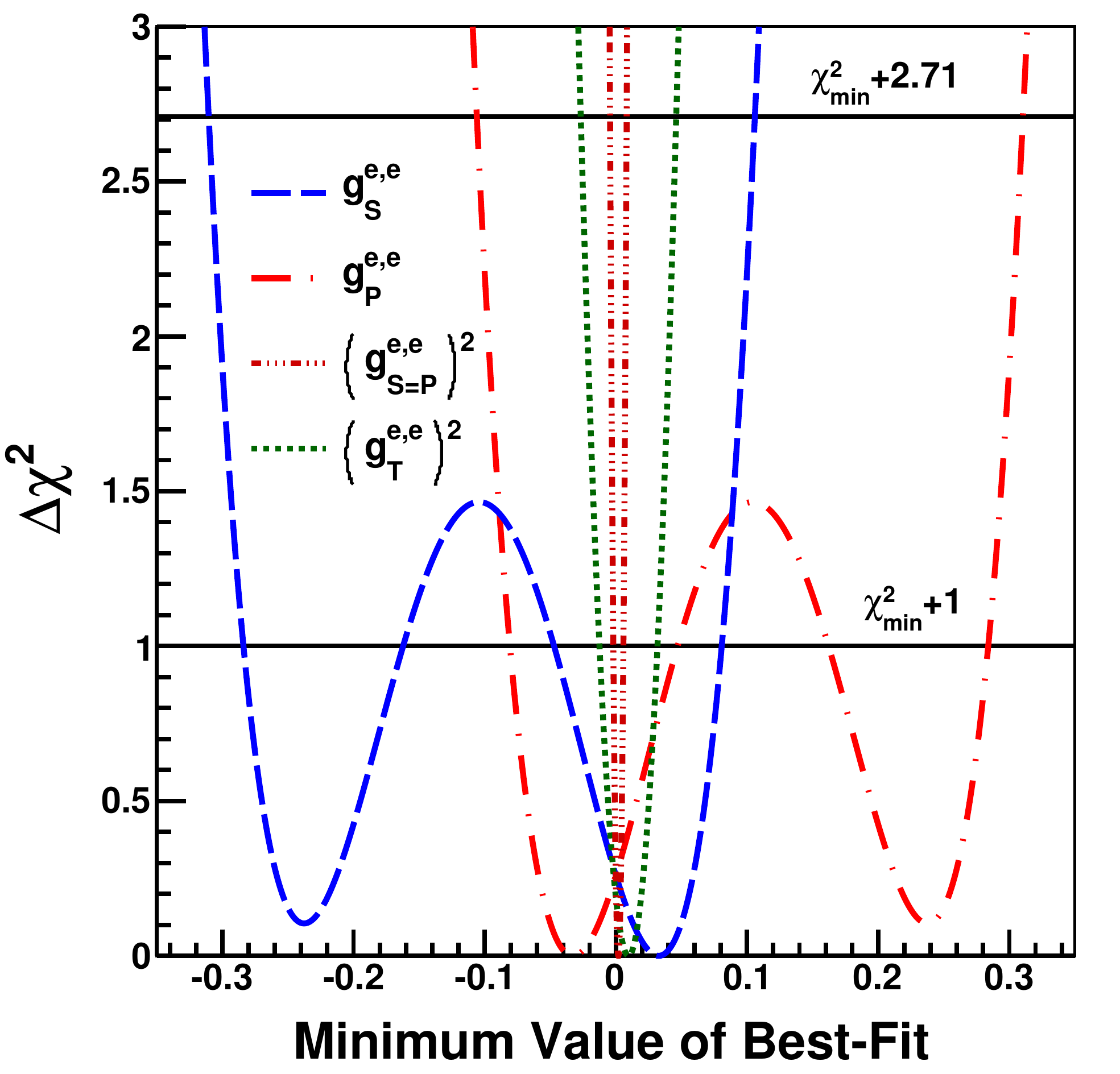} \qquad
  \includegraphics[width=8cm]{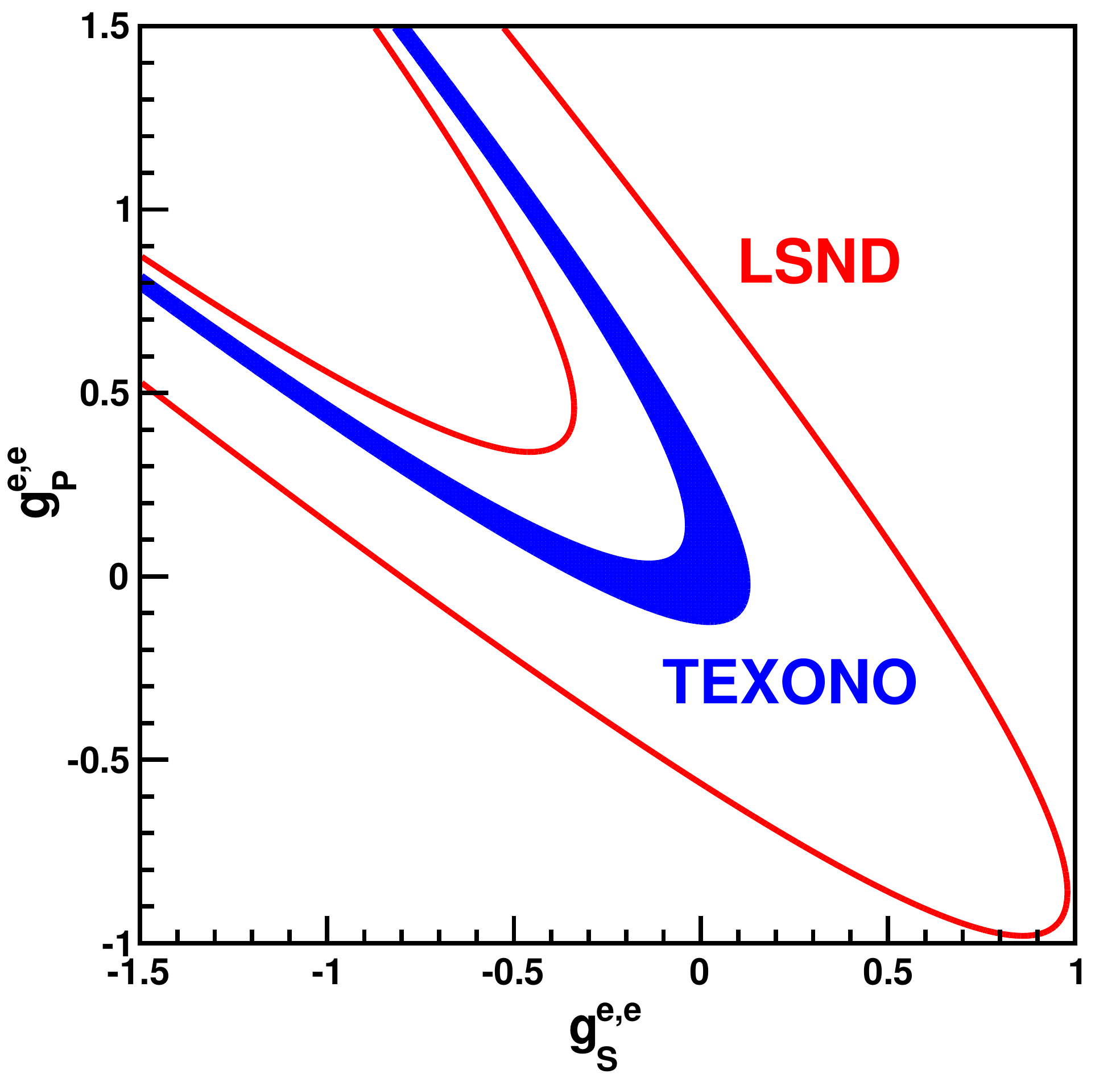} \\
{\bf \hspace{0.1cm}(c)} {\bf \hspace{8.5cm}(d)} \\
  \includegraphics[width=8cm]{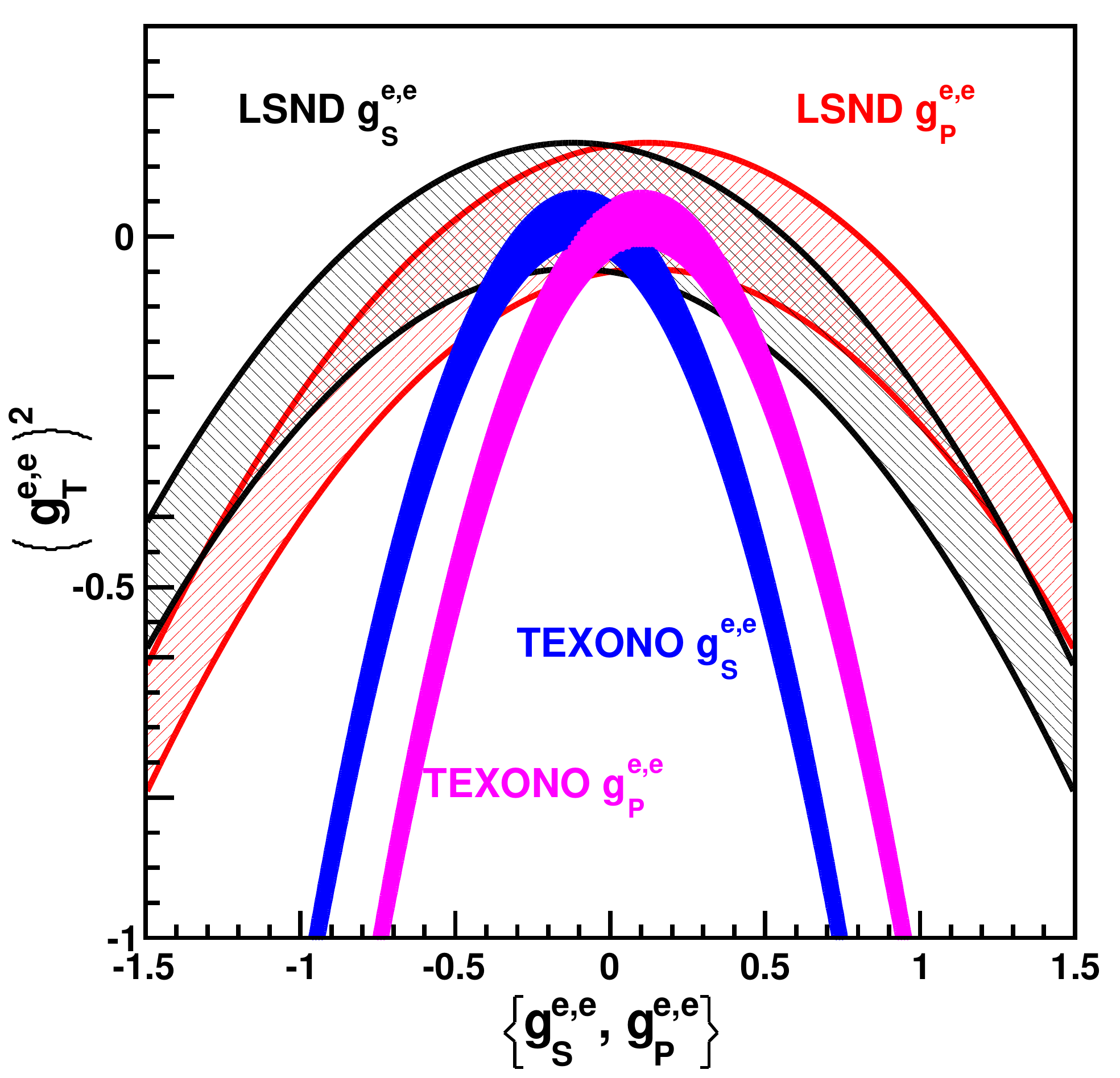}\qquad
  \includegraphics[width=8cm]{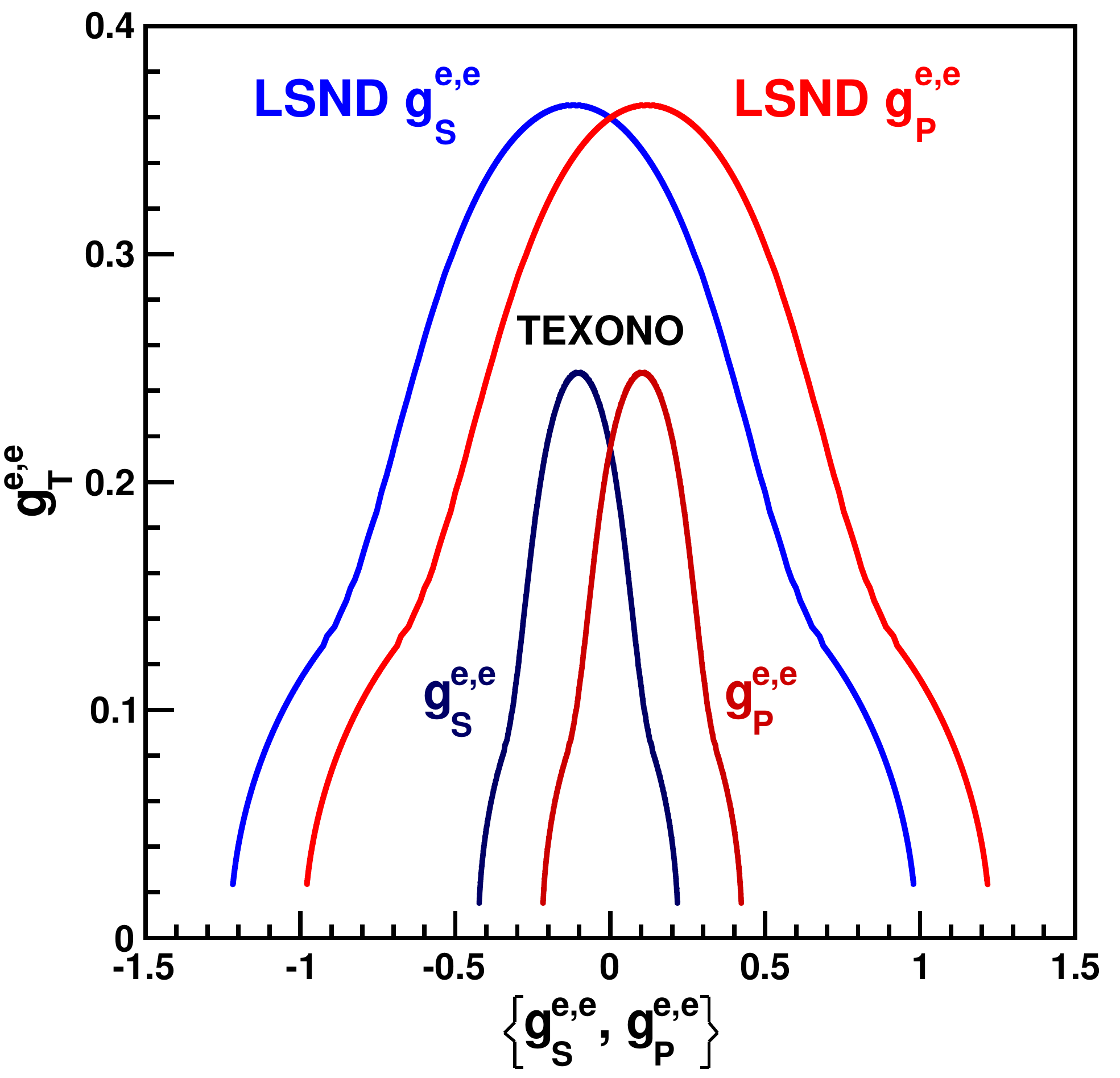}
  \caption{
  (a) $\Delta\chi^{2}$ of one-parameter-at-a-time analysis for
  $g_{S,P}^{ee}$, $\left(g_{S=P}^{e,e}\right)^2$, and
  $\left(g_{T}^{e,e}\right)^2$.
  The allowed region at 90\% C.L. for TEXONO and LSND experiments
  in (b) $g_{S}^{e,e}-g_{P}^{e,e}$,
  (c) $g_{S}^{e,e}-{\left(g_{T}^{e,e}\right)}^2 $ and
  $ g_{P}^{e,e}-{\left(g_{T}^{e,e}\right)}^2 $
  parameter spaces.
  (d) Upper limits at 90\% C.L. of $g_{T}^{e,e}$ as
  functions of $g_{S}^{e,e}$ and $g_{P}^{e,e}$.
  }
  \label{fig::gs_gp_gt_90cl}
  \end{center}
\end{figure*}

Observing NSIs would imply the existence of right-handed
neutrinos; therefore, it is an important channel for studying
new physics BSM. However, there are few studies
that exist on scalar-, pseudoscalar-, or tensorial-type NSIs in
the literature, mainly due to the motivation of Vector-Axialvector 
(V-A) structure of the SM and the assumption of their small 
contributions to the cross section. To overcome this deficiency, 
in addition to FC and FV NSI parameters scalar, pseudoscalar, 
and tensorial NSIs of neutrinos are studied via the (anti)neutrino-electron 
interaction channel and new limits are set to the related parameters 
by adopting a model-independent method introduced in this paper.

The effective Lagrangian for scalar  and/or pseudoscalar
interaction~\cite{gaitan} can be written as
\begin{equation}
\mathcal{L}_{S,P} = \sum_{\alpha}\sum_{\beta}\overline{l}_{\alpha}\left(
\mathcal{O}_{S,P} \right)
\nu_{\beta}~,
\label{eq::L_leptons-H}
\end{equation}
where $\left(\mathcal{O}_{S,P} \right)$ is a general operator with
scalar/pseudoscalar interactions.

The effective Lagrangian for tensorial NSI~\cite{barranco2012}
interaction can be written as
\begin{eqnarray}
-\mathscr{L}_\text{T}^\text{eff} = \varepsilon_{\alpha\beta}^{fT}2\sqrt{2}G_F
(\bar{\nu}_\alpha \sigma^{\mu\nu} \nu_\beta) (\bar{f}\sigma_{\mu\nu}f)
\label{eq::tNSI_lag}
\end{eqnarray}
with $\sigma^{\mu\nu} = [\gamma^{\mu},\gamma^{\nu}]
= \gamma^{\mu}\gamma^{\nu} - \gamma^{\nu}\gamma^{\mu}$
and $\alpha,\beta=\,e,\,\mu,\,\tau$.

The differential cross section of scalar-pseudoscalar NSI
for $\nu_e - e^- $ and $ \bar{\nu}_e - e^-$,
respectively, can be written as
\begin{eqnarray}
\left[ \frac{d\sigma _{\nu_e,e}}{dT}\right]_\text{S,P}^\text{NSI}
&=&\frac{2G_{F}^{2}m_{e}}{\pi}
\left\{\left[\left( \left\vert g^{e,e}_{S}\right\vert
+\left\vert g^{e,e}_{P}\right\vert \right) ^{2} \right. \right. \nonumber \\
&+& \left.g_{R}\textrm{Re}\left(
g^{e,e}_{S}-g^{e,e}_{P}\right)^{\textcolor{white}{2}}\hspace{-0.15cm}\right]
\left( 1-\frac{T}{E_{\nu}}\right)^{2} \nonumber \\
&-& \left.\left(g_{L}+1\right)\textrm{Re}\left( g^{e,e}_{S}-g^{e,e}_{P}\right)
 \frac{m_{e}T}{2E_{\nu}^{2}} \right\}
\label{eq::sps_v-e}
\end{eqnarray}
and
\begin{eqnarray}
\left[ \frac{d\sigma_{\bar{\nu}_e,e}}{dT}\right]_\text{S,P}^\text{NSI}
&=&\frac{2G_{F}^{2}m_{e}}{\pi}
\left\{\left( \left\vert g^{e,e}_{S}\right\vert
+\left\vert g^{e,e}_{P}\right\vert \right) ^{2} \right. \nonumber \\
&+& g_{R}\textrm{Re}\left(
g^{e,e}_{S}-g^{e,e}_{P}\right)\nonumber \\
&-& \left.\left(g_{L}+1\right)\textrm{Re}\left( g^{e,e}_{S}-g^{e,e}_{P}\right)
 \frac{m_{e}T}{2E_{\nu}^{2}} \right\}~.
\label{eq::sps_antiv-e}
\end{eqnarray}

In the tensorial NSI case, there will be no interference between
SM interaction channels of neutral and charged currents since
incoming and outgoing neutrinos' helicities would be different
at the exchange vertex. In that case the contribution of tensorial
interaction to the SM cross section should be just added numerically.
The differential cross section of tensorial NSI can be
written as~\cite{bkayser87,barranco2012}
\begin{eqnarray}
\left[ \frac{d\sigma}{dT}\right] _\text{T}^\text{NSI}
= \frac{2G_{F}^{2}m_{e}}{\pi} \sum\limits_{\beta = e, \mu, \tau}
\left({\varepsilon^{eT}_{e\beta}}\right)^2
\left[2{\left(1-\frac{T}{2E_{\nu }}\right)}^2\right.
\nonumber\\
-\left.\frac{m_eT}{2E_\nu^2}\right]~,
\label{eq::tNSI}
\end{eqnarray}
where $\varepsilon^{eT}_{e\beta}$ is the strength of the tensorial
NSI coupling on electrons.

Since we will only consider one parameter
at a time in the analysis and the contribution of $\varepsilon^{eT}_{ee}$
is the same as $\varepsilon^{eT}_{e\beta}$ for $\beta\neq e$ case,
we will denote $\varepsilon^{eT}_{e\beta}$ as $g_{T}^{e,e}$ throughout
the paper, the same as in the literature.

The measurable recoil spectra at a typical reactor flux of
$\phi({\bar{\nu}_e}) = 10^{13}~ \text{cm}^{-2}\text{s}^{-1}$
are displayed in Fig.~\ref{fig::diff_nsi} for typical scalar,
pseudoscalar and tensorial NSI parameters.
The spectral shapes of NSI contributions for all types give rise 
to quite similar to the SM one. Accordingly, like FC and FV NSIs, 
the most suitable energy range to study scalar, pseudoscalar, and 
tensorial NSIs is in MeV energy range where the SM effects were 
measured with good accuracy.

Like the FC and FV NSI parameters, the scalar, pseudoscalar and 
tensorial NSI parameters~\cite{barranco2012,kjh2013,dkptsk}
are also constrained by the accuracy of the SM cross section
measurements. Accordingly, only CsI(Tl) data set for TEXONO
and LSND experiments are adopted for scalar, pseudoscalar and
tensorial NSI analysis.

The scalar and pseudoscalar NSI parameters $g_{S,P}^{e,e}$
given in \Eq(\ref{eq::sps_v-e}) and \Eq(\ref{eq::sps_antiv-e})
are the fitting variables in the minimum-$\chi^2$ analysis.
If $g_{S}^{e,e}=g_{P}^{e,e}$ then there will be some
simplifications in these equations. In this case
${\left(g^{e,e}_{S=P}\right)}^2$ becomes the fitting parameter.

By adopting one-parameter-at-a-time analysis, from the best fit,

\begin{eqnarray}
g_{S}^{e,e} &=& \left[3.27 \pm 6.39 \pm 3.10 \right]\times10^{-2}~, \nonumber \\
g_{P}^{e,e} &=& \left[-3.27 \pm 6.39 \pm 3.10 \right]\times10^{-2}~,
\nonumber \\
{\left(g^{e,e}_{S=P}\right)}^2 &=& \left[0.19 \pm 0.38 \pm 0.31 \right]\times10^{-2}
~~\text{, and} \nonumber \\
{\left(g_{T}^{e,e}\right)}^2 &=&
\left[0.96 \pm 2.21 \pm 1.82\right]\times10^{-2}
\label{eq::gsgpgt1par_tex}
\end{eqnarray}
are obtained at $\chi^2_{min} / \text{dof} = 8.7/9$.

These results are converted to the bounds for the scalar and pseudoscalar NSIs
but only  upper limits for tensorial and $|g_{S=P}^{e,e}|$ NSI couplings as
\begin{eqnarray}
-0.317 < g_{S}^{e,e} < 0.113 ~~ (\text{for}~~ g_{P}^{e,e}=0)~~, \nonumber \\
-0.113 < g_{P}^{e,e} < 0.317 ~~ (\text{for}~~ g_{S}^{e,e}=0)~~,  \nonumber \\
|g_{S=P}^{e,e}| < 0.100 ~~ (\text{for}~~ g_{S}^{e,e}=g_{P}^{e,e}) ~~\text{, and} \nonumber \\
g_{T}^{e,e} < 0.238 ~~ (\text{for}~~ g_{S}^{e,e}=g_{P}^{e,e}=0)
\label{eq::gsgpgt_tex_bound}
\end{eqnarray}
at 90 \% C.L. for TEXONO CsI(Tl) data set.

The best-fit results and $\chi^2$ behaviors
of the scalar, pseudoscalar, and tensorial NSI parameters of
$g_{S,P}^{ee}$, $\left(g_{S=P}^{e,e}\right)^2$ and $\left(g_{T}^{e,e}\right)^2$
which are adopted as fitting variables are illustrated in
Fig.~\ref{fig::gs_gp_gt_90cl}(a).

Similarly,
\begin{eqnarray}
g_{S}^{e,e} &=& 0.27 \pm 0.59 \pm 0.26, \nonumber \\
g_{P}^{e,e} &=& -0.27 \pm 0.59 \pm 0.26, \nonumber \\
{\left(g^{e,e}_{S=P}\right)}^2 &=& \left[3.47 \pm 4.78 \pm 4.36 \right]\times10^{-2}
~~\text{, and} \nonumber \\
{\left(g_{T}^{e,e}\right)}^2 &=&
\left[3.96 \pm 5.47 \pm 4.97 \right]\times10^{-2}
\label{eq::gsgpgt1par_lsnd}
\end{eqnarray}
are derived for the LSND experiment, and they correspond to
the limits of
\begin{eqnarray}
-0.880<g_{S}^{e,e}<0.642 ~~ (\text{for}~~ g_{S}^{e,e}=0),  \nonumber \\
-0.642 <g_{P}^{e,e}<0.880 ~~ (\text{for}~~ g_{P}^{e,e}=0), \nonumber \\
|g_{S=P}^{e,e}| < 0.375 ~~ (\text{for}~~ g_{S}^{e,e}=g_{P}^{e,e})
~~\text{, and} \nonumber \\
g_{T}^{e,e} < 0.401 ~~ (\text{for}~~ g_{S}^{e,e}=g_{P}^{e,e}=0)
\label{eq::gsgpgt_lsnd_bound}
\end{eqnarray}
at 90 \% C.L. which are similar to published results
of LAMPF experiment reported in Ref.~\cite{allen93} since
their experimental sensitivities are similar.

The results of one-parameter-at-a-time analysis
for TEXONO and LSND experiments are listed in Table~\ref{tab::gs_gp}.
As can be seen, TEXONO CsI(Tl) data set provides much better
constraints than those from LSND or LAMPF accelerator experiments
and also from the other reactor neutrino experiment data sets
given in Ref.~\cite{barranco2012}.

The allowed regions at 90\% C.L. from two-parameter analysis
in the [$g_{S,P}^{e,e}$ and ${\left(g_{T}^{e,e}\right)}^2]$
parameter spaces are displayed in Fig.~\ref{fig::gs_gp_gt_90cl}
in which LSND results are superimposed for complementarity.
The 90\% CL upper limits of $g_{T}^{e,e}$ as functions of
$g_{S}^{e,e}$ and $g_{P}^{e,e}$ for the two experiments
are displayed in Fig.~\ref{fig::gs_gp_gt_90cl}(d).
It can be seen that the TEXONO data provide more stringent
constraints than those from LSND  in  both the
$g_{S}^{e,e} - g_{P}^{e,e}$  and $g_{S,P}^{e,e} - g_{T}^{e,e}$
parameter spaces.
\begin{figure}[!ht]
  \begin{center}
  \includegraphics[width=8cm]{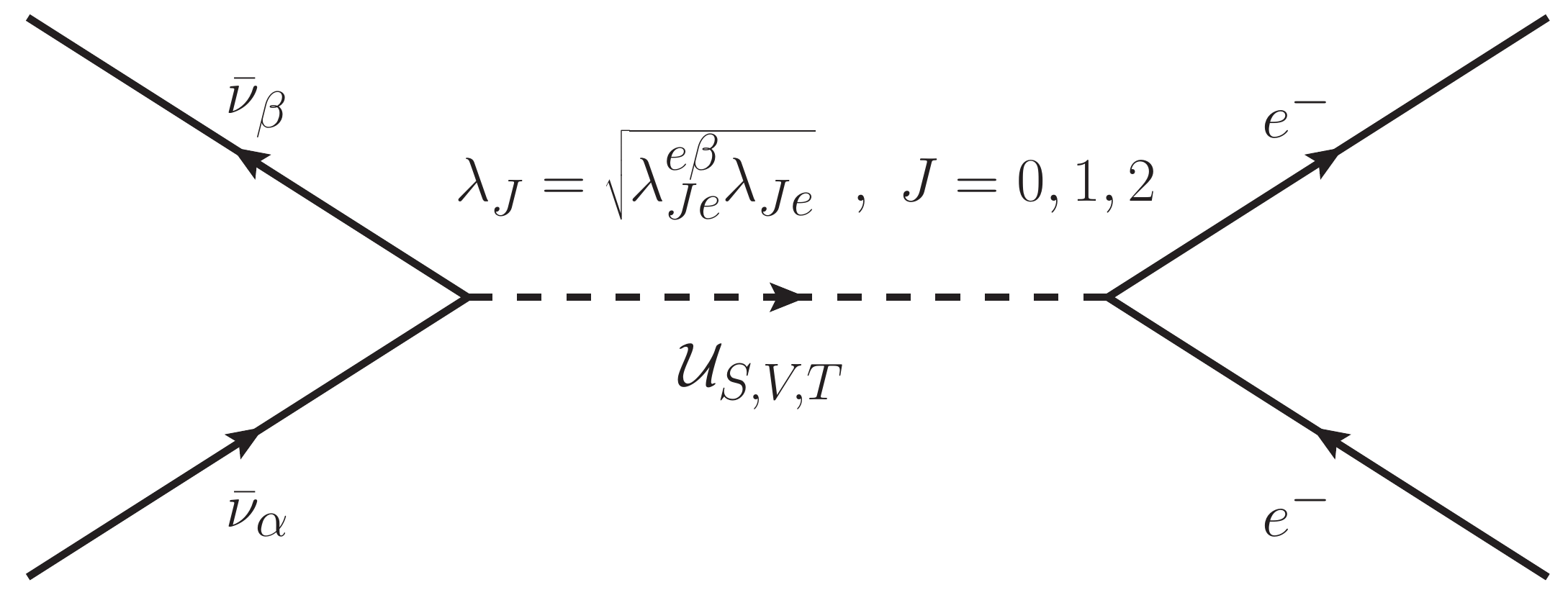}
  \caption{Interactions of a neutrino with an electron via exchange
  of massive mediators such as a virtual unparticle
  (scalar ${\cal U_S}$, vector ${\cal U_V}$, or tensorial $\cal U_T$).}
  \label{fig::feydiag_up}
  \end{center}
\end{figure}
\begin{figure}[!ht]
  \begin{center}
  \includegraphics[width=8cm]{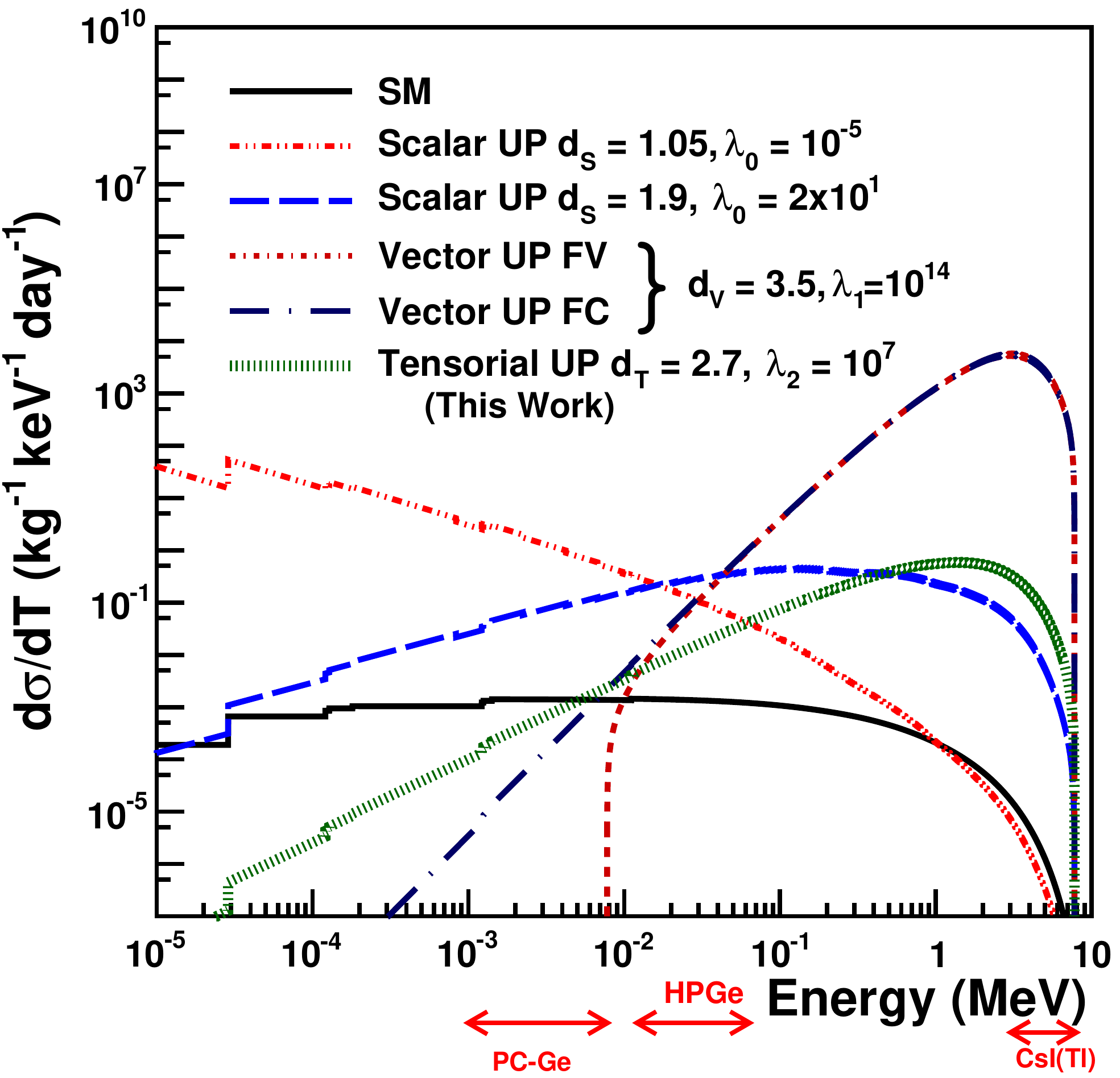}
  \caption{Differential cross section as a function of
  the recoil energy $T$ with typical reactor $\nuebar$ spectra
  for scalar UPs at two values of $( \dsca , \lambda_{0} )$,
  for vector UPs at a value of $( \dvec , \lambda_{1} )$ for both
  FV and FC UP cases~\cite{mdeniz033004},
  and for tensorial UPs at a value of $(d_T, \lambda_2)$ (this work).
  The relevant energy ranges of the three
  data sets used in the present analysis are also shown.}
  \label{fig::diff_up}
  \end{center}
\end{figure}
\begin{figure*}[!ht]
  \begin{center}
  {\bf \hspace{0.5cm}(a)} {\bf \hspace{8.5cm}(b)}\\
    \includegraphics[width=8cm]{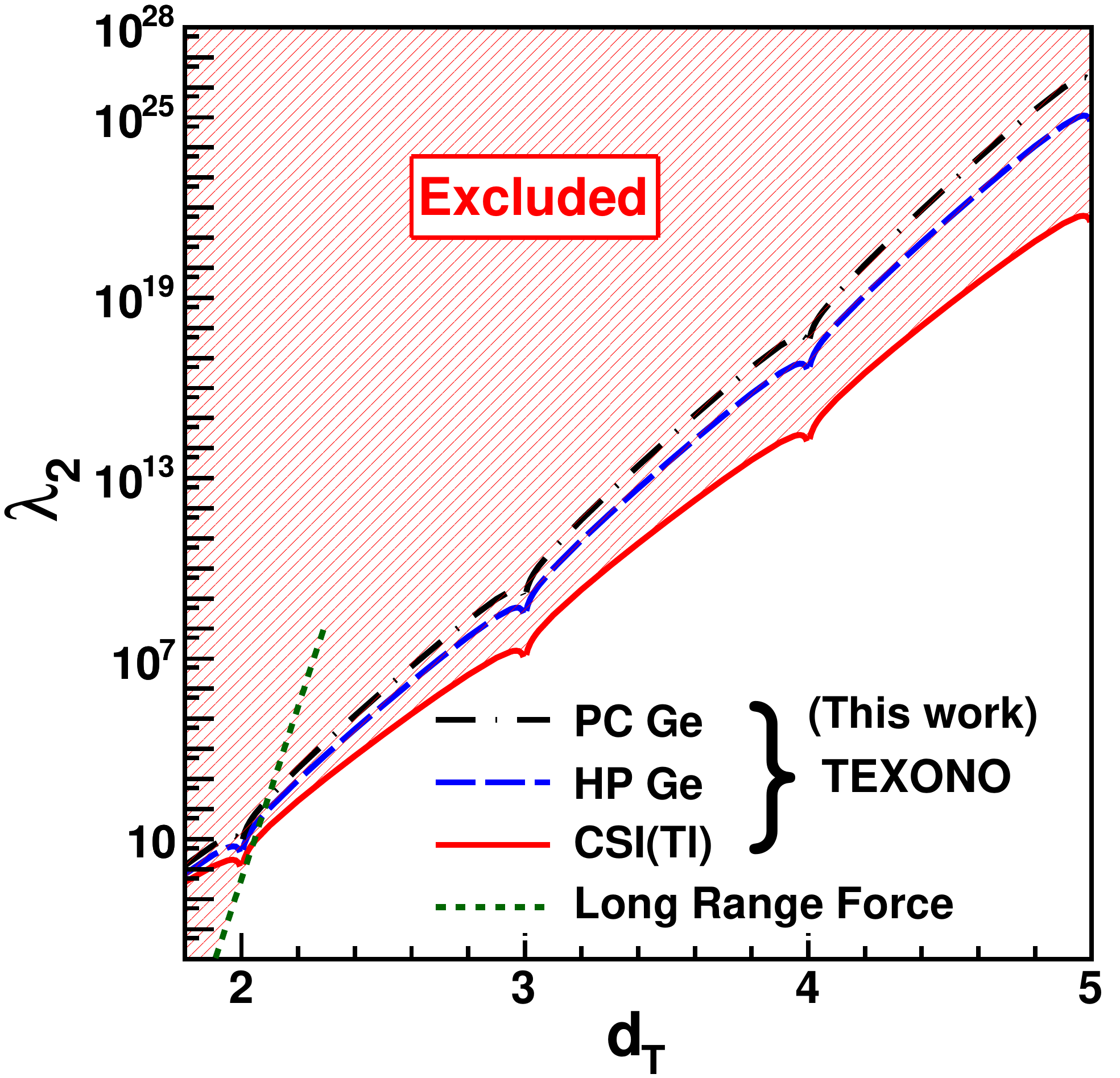}\qquad
    \includegraphics[width=8cm]{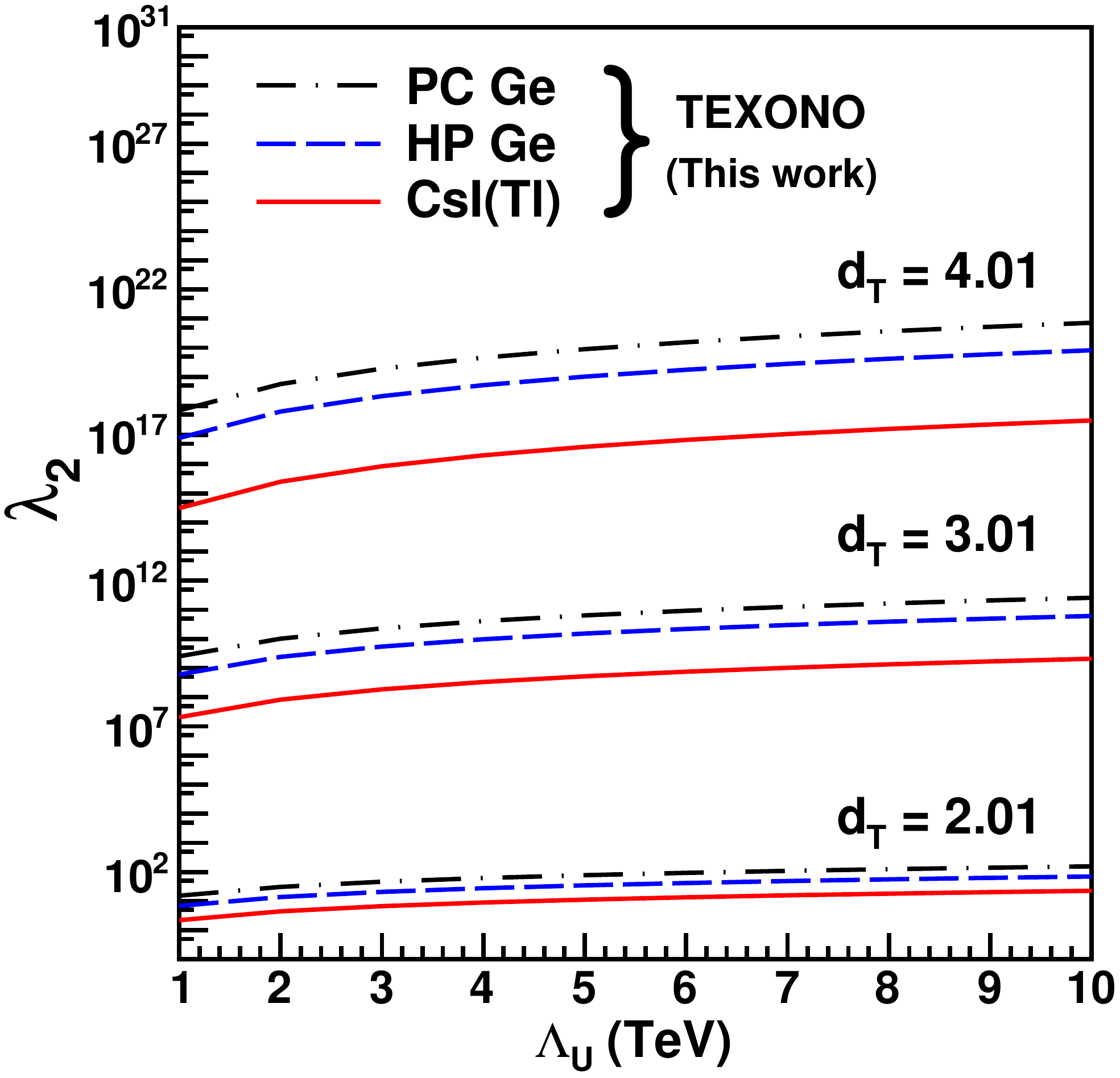}
  \caption{Constraints on UPs with tensorial
  exchange (a) The coupling $\lambda_2$ vs mass dimension
  $d_T$ at $\Lambda_U=1$ TeV for the three data sets adopted for
  this analysis in which the long-range force result is superimposed for comparison.
  (b) Upper bounds on $\lambda_2$ at different energy scales $\Lambda_U$.
  The  space above the lines is excluded.}
  \label{fig::up_90}
  \end{center}
\end{figure*}

\section{Unparticle Physics}

Unparticle physics was first presented by Georgi with two articles in 2007. 
Besides SM fields, a sector is assumed to be scale invariant at high energies. 
These fields are called Banks-Zaks (${\cal BZ}$) fields. In this model SM 
fields interact with ${\cal BZ}$ fields by an exchange of mass scale-invariant 
massive particles that are called unparticles~\cite{Georgi2007,Lenz2007}.

The effective Lagrangian is given by~\cite{barranco2012,Bolanos2007}
\begin{eqnarray}
{\cal L_U} = C_{\cal O_U} \frac{\Lambda_{\cal U}^{d_{\cal BZ}-d_{\cal U}}}
{M_{\cal U}^{d_{\cal SM}+d_{\cal BZ}-4}} {\cal O_{SM}}{\cal O_U}~,
\label{eq::lup}
\end{eqnarray}
where ${\cal O_U}$ is the unparticle operator of scaling dimension $d_{\cal U}$
in the low-energy limit and $C_{\cal O_U}$ is the dimensionless coupling constant.
The unparticle operator can be a scalar, vector, spinor, or tensor type.

The unparticle can directly be studied in accelerator experiments via 
investigating missing energy signals in the detection channel~\cite{cheung2007}, 
but alternatively its effect can also be examined in the neutrino-electron 
elastic scattering channel as being a virtual mediator particle. The latter 
approach is adopted in this analysis using reactor neutrinos as probes.
Scalar and vector UPs for both FV and FC cases were studied in our early 
work~\cite{mdeniz033004}; in this paper we focus on tensorial UP interaction.

The interaction Lagrangian for $\nu_{\alpha} + e \rightarrow \nu_{\beta} + e$
depicted in Fig.~\ref{fig::feydiag_up} via tensorial unparticle exchange 
is given by ~\cite{upchen07,upbalantekin07,upbarranco09,eagarces,cheung2007,hur2007}
\begin{eqnarray}
\mathscr{L}_\text{J=2} & = &
\frac{-i}{4} \frac{\lambda_{2e}}{\Lambda_{\cal U}^{d_{\cal T}}} \bar{e}
\left(\gamma_\mu \overleftrightarrow{D}_\nu + \gamma_\nu
\overleftrightarrow{D}_\mu\right)\psi_e{\cal O}^{\mu\nu}_{\cal U}\nonumber \\
& + & \frac{\lambda_{2\nu}^{\alpha\beta}}{\Lambda_{\cal U}^{d_{\cal T}}}
{\cal F}_{\mu\alpha} {\cal F}_\nu^\alpha {\cal O}^{\mu\nu}_{\cal U}~,
\label{eq::UP_ten_lag}
\end{eqnarray}
where ${\cal F}_{\mu\nu}$ is the gauge field strength and
$\lambda_{2 e}$ and $\lambda_{2 \nu}^{\alpha\beta}$ are
the corresponding coupling constants.

The cross section of $\bar{\nu}_e - e$ scattering via
tensorial UP exchange is given by~\cite{barranco2012}
\begin{eqnarray}
{\left(\frac{d \sigma}{d T}\right)}_{\cal U_T} &=& ~\frac{f^2(d_T)}{\pi\Lambda_{\cal U}^{4d_T-4}}
~2^{2d_T-3}~m_e^{2d_T-3}~T^{2d_T-4}\nonumber\\
&\times& \left[3\left(1-\frac{T}{2E_\nu} \right)^2
-\frac{m_eT}{2E_\nu^2}\right]~,
\label{eq::UP_ten}
\end{eqnarray}
where
\begin{eqnarray}
f(d_X) = \frac{\lambda_X^2}{2\sin{(d_X\pi)}}A(d_X)
\label{eq::f}
\end{eqnarray}
and the normalization constant $A(d_X)$ is given by
\begin{eqnarray}
A(d_X) = \frac{16\pi^{5/2}}{{(2\pi)}^{2d_X}}
\frac{\Gamma(d_X+1/2)}{\Gamma(d_X-1)\Gamma(2d_X)}~.
\label{eq::a}
\end{eqnarray}

The cross section of $\nuebare$ scattering via all kinds of UP exchange
can be obtained by making a replacement of $d_X \rightarrow d_S, d_V $,
or $d_T$ and $ \lambda_X \rightarrow \lambda_0, \lambda_1$, or $\lambda_2$
representing scalar, vector, and tensorial UP interactions, respectively.

The differential cross sections of scalar, both FC and FV vector- and 
tensor-type UP interactions at a typical mass dimension of $d$ using TEXONO 
CsI(Tl), HP-Ge, and PC-Ge detector data sets are displayed in 
Fig.~\ref{fig::diff_up}, where the SM contribution is superimposed for 
comparison. The sawtooth structures for $T \alt 1 ~{\rm keV}$ are due to 
suppression by the atomic binding energy~\cite{jiunn011301,atombinding}. 
As illustrated in the figure the cross sections of different UP type give 
different behavior with respect to the recoil energy.

As are the cases of both FC and FV vector UPs,  
studies in the high energy regime offer greater advantage 
than those at low energy for both tensorial UPs and scalar UPs
with higher mass dimension of $d$.  On the other hand, the 
low-energy regime is more favorable for scalar UPs with low mass 
dimension of $d$. Since the cross sections of the SM are measured 
more precisely in the MeV energy range, more sensitive results are expected 
for tensorial UPs with the CsI(Tl) detector data set compared to those from 
Ge detector data sets. Since different ranges of $d_T$ for all the data 
sets listed above give different and comparable sensitivities, all the 
three data sets of $\nuebare$ are used in the tensorial UP
analysis for their complementarity.

Three parameters, unparticle mass dimension $d_T$, unparticle energy scale
$\Lambda_U$  and coupling constant
$ \lambda_2 \equiv \sqrt{ \lambda ^ {e\beta}_{2\nu} \lambda_{2e}}$
characterizing the unparticle interactions can be probed experimentally.
There is a bound on $d_T$ as $d_T \geq 2$ for the antisymmetric tensor and
$d_T \geq 4$ for the symmetric case~\cite{grinstein08}.
The UP energy scale is taken to be $\Lambdaup \sim 1~ {\rm TeV}$ as in most
recent works~\cite{upbalantekin07,upmontanino08,upbarranco09} as well as
$\Lambdaup$ up to 10~ {\rm TeV}.

Constraints on $\lambda_2$ at different $d_T$ in the case of tensorial UP
exchange interaction are derived at $\Lambda_U=1$ TeV. The results are shown
in Fig.~\ref{fig::up_90}(a), in which a long-range force result
reported in Ref.~\cite{upbarranco09} is superimposed for comparison.

To observe the UP signal from the data and to set bounds on it, among
the investigations of the collider, CP-violation, deep-inelastic, lepton
flavor-violating, hadron mixing and decay, neutrino interaction, nucleon
decay experiments, and cosmological and astroparticle efforts, there is
another unparticle approach that involves long-range forces operating
at macroscopic distances and governed by a nonintegral power law, which
was first introduced in Ref.~\cite{Liao07}. It is based on spin-dependent
interactions between electrons and related to the Newtonian gravitational
inverse squared law mediated by a tensorlike ungravity 
interaction~\cite{Gonzales08}.

The upper bounds for $\lambda_2$ at different energy
scale $\Lambda_U$ up to 10 TeV are shown in Fig.~\ref{fig::up_90}(b).
As can be seen, the limits are improved by the TEXONO CsI(Tl) data set
for $d_T>2.04$, and CsI(Tl) gives rise to more stringent limits 
at larger $d_T$.

Since ${(d\sigma/dT)}_{\cal U_T}$ is proportional to $\lambda_2^4$,
which can be seen from \Eq(\ref{eq::UP_ten}), the potential of placing more
severe constraints on the coupling constants due to improved experimental
sensitivities can only be modest. When the measurements of CsI(Tl) or sub-keV Ge 
can achieve ~1\% precision, an improvement by a factor of 3 can be expected for 
the sensitivity of $\lambda_2$~\cite{mdeniz033004, sbilmis073011,nuNcohsca}.

\section{SUMMARY AND PROSPECTS}

In summary of this article, some of the model-independent
BSM new physics scenarios such as scalar, pseudoscalar, and 
tensorial NSIs and the tensorial component of unparticle physics 
are discussed. The investigation on the neutrino oscillation phenomena 
enters the high precision era. Therefore, the existence and contribution
of NSIs would help to increase the precision of the experiments.

The experimental results of upper bounds for NSIs using data from
the analysis of $\nuebare$ and $\nuee$ elastic scattering interaction
cross section measurements are placed in the framework of these BSM
scenarios and the existing experimental sensitivities are improved.
We have found new constraints on the relevant parameters
that are more stringent than previous laboratory constraints.

Indeed, TEXONO provides better sensitivity in NSI parameter space
compared to the LSND experiment; therefore, combined results will
improve the existing bounds. Moreover, with this study we
extend the parameter space as well by covering the low-mass
dimension of $d$ in UP physics studies.

A new research avenue to investigate  BSM theories can be opened via the 
studies of $\nuebar$-nucleus coherent scattering. This is one of the 
themes of the on-going TEXONO low-energy neutrino physics programs.


\section{ACKNOWLEDGMENTS}

This work is supported by contract 114F374 under Scientific and Technological 
Research Council of Turkey (T\"{U}B\.ITAK);
104-2112-M-001-038-MY3 from the Ministry of Science and Technology, Taiwan and 
2015-ECP4 from the National Center of Theoretical Sciences, Taiwan.


\begin{thebibliography}{99}
%
\bibitem{king2015}
S.F.~King, \href{http://iopscience.iop.org/0954-3899/42/12/123001}
{J. Phys. G {\bf 42}, 123001 (2015)}.

\bibitem{jpanman-wjmarciano}
J.~Panman, {\it Precision Tests of the Standard Electroweak Model},
edited by P.~Langacker (World Scientific, Singapore, 1995), p. 504;
W.J.~Marciano and Z.~Parsa, \href{http://dx.doi.org/10.1088/0954-3899/29/11/013}
{J. Phys. G {\bf 29}, 2629 (2003)}.

\bibitem{ohlsson2013}
T.~Nilsson, \href{http://dx.doi.org/10.1088/0034-4885/76/4/044201}
{Rep. Prog. in Phys. {\bf 76}, 044201 (2013)}.

\bibitem{miranda2015}
O.G.~Miranda and H.~Nunokawa, \href{http://iopscience.iop.org/1367-2630/17/9/095002}
{New J. Phys. {\bf 17}, 095002 (2015)}.

\bibitem{jerler125}
J.~Erler and P.~Langacker, \href{http://dx.doi.org/10.1016/j.physletb.2008.07.027}
{Phys. Lett. B {\bf 667}, 125 (2008)}, and references therein.

\bibitem{mdeniz033004}
M.~Deniz \textit{et al.}, \href{http://dx.doi.org/10.1103/PhysRevD.82.033004}
{Phys. Rev. D {\bf 82}, 033004 (2010)}.
 
\bibitem{sbilmis073011}
S.~Bilmi\c{s} \textit{et al.}, \href{http://dx.doi.org/10.1103/PhysRevD.85.073011}
{Phys. Rev. D {\bf 85}, 073011 (2012)}.

\bibitem{jiunn011301}
J.W.~Chen, H.C.~Chi, H.B.~Li, C.P.~Liu, L.~Singh, H.T.~Wong,
C.L.~Wu, and C.P.Wu, \href{http://dx.doi.org/10.1103/PhysRevD.90.011301}
{Phys. Rev. D {\bf 90}, 011301(R) (2014)}.

\bibitem{sbilmis033009}
S.~Bilmis, I.~Turan, T.M.~Aliev, M.~Deniz, L.~Singh, and H.T.~Wong,
\href{http://dx.doi.org/10.1103/PhysRevD.92.033009} {Phys. Rev. D {\bf 92}, 033009 (2015)}.

\bibitem{bkayser87}
B.~Kayser,  E.~Fischbach, S.P.~Rosen, and H.~Spivack, \href{http://dx.doi.org/10.1103/PhysRevD.20.87}
{Phys. Rev. D {\bf 20}, 87 (1979)}.

\bibitem{mdeniz072001}
M.~Deniz \textit{et al.}, \href{http://dx.doi.org/10.1103/PhysRevD.81.072001}
{Phys. Rev. D {\bf 81}, 072001 (2010)}.

\bibitem{hbli-htwong}
H.B.~Li \textit{et al.}, \href{http://dx.doi.org/10.1103/PhysRevLett.90.131802}
{Phys. Rev. Lett. {\bf 90}, 131802 (2003)};
H.T.~Wong \textit{et al.}, \href{http://dx.doi.org/10.1103/PhysRevD.75.012001}
{Phys. Rev. D {\bf 75}, 012001 (2007)};
H. M. Chang et al., Phys. Rev. D {\bf 75}, 052004 (2007) 
{http://dx.doi.org/10.1103/PhysRevD.75.052004}.

\bibitem{soma2016}
A.K.~Soma \textit{et al.}, \href{http://dx.doi.org/10.1016/j.nima.2016.08.044}
{Nucl. Instrum. Methods Phys. Res., Sect. A {\bf 836}, 67 (2016)}.

\bibitem{auerbach2001}
L.B.~Auerbach \textit{et al.}, \href{http://dx.doi.org/10.1103/PhysRevD.63.112001}
{Phys. Rev. D {\bf 63}, 112001 (2001)}.

\bibitem{Gary98}
G.J. Feldman and R.D. Cousins, \href{http://dx.doi.org/10.1103/PhysRevD.57.3873}
{Phys. Rev. D {\bf 57}, 3873 (1998)}.

\bibitem{nsiastrophys}
N.~Fornengo, M.~Maltoni, R.T.~Bayo and J.W.F.~Valle, \href{http://dx.doi.org/10.1103/PhysRevD.65.013010}
{Phys. Rev. D {\bf 65}, 013010 (2001)};
P.S.~Amanik, G.M.~Fuller and B.~Grinstein, \href{http://dx.doi.org/10.1016/j.astropartphys.2005.06.004}
{Astropart. Phys.  {\bf 24}, 160 (2005)};
G.L.~Fogli, E.~Lisi, A.~Mirizzi and D.~Montanino, \href{http://dx.doi.org/10.1103/PhysRevD.66.013009}
{Phys. Rev. D {\bf 66}, 013009 (2002)};
A.~Esteban-Pretel, R.~Tomas and J.W.F.~Valle, \href{http://dx.doi.org/10.1103/PhysRevD.76.053001}
{Phys. Rev. D {\bf 76}, 053001 (2007)}.

\bibitem{reactornubsm}
O.G.~Miranda, M.~Maya and R.~Huerta, \href{http://dx.doi.org/10.1103/PhysRevD.53.1719}
{Phys. Rev. D {\bf 53}, 1719 (1996)};
O.G.~Miranda, V.~Semikoz and J.W.F.~Valle, \href{http://dx.doi.org/10.1016/S0920-5632(98)00050-4}
{Nucl. Phys. Proc. Suppl. {\bf 66}, 261 (1998)};
J.~Barranco, O.G.~Miranda and T.I.~Rashba, \href{http://dx.doi.org/10.1103/PhysRevD.76.073008}
{Phys. Rev. D {\bf 76}, 073008 (2007)};
A.~Bolanos \textit{et al.}, \href{http://dx.doi.org/10.1103/PhysRevD.79.113012}
{Phys. Rev. D {\bf 79}, 113012 (2009)}.

\bibitem{nsiboundlsnd}
S.~Davidson \textit{et al.}, \href{http://dx.doi.org/10.1088/1126-6708/2003/03/011}
{J. High Energy Phys. {\bf 03} (2003), 011}.

\bibitem{nsiboundcombined}
J.~Barranco \textit{et al.}, \href{http://dx.doi.org/10.1103/PhysRevD.73.113001}
{Phys. Rev. D {\bf 73}, 113001 (2006)};
J.~Barranco \textit{et al.}, \href{http://dx.doi.org/10.1103/PhysRevD.77.093014}
{Phys. Rev. D {\bf 77}, 093014 (2008)}.

\bibitem{cbig}
C.~Biggio, M.~Blennow, and E.~Fernandez-Martinez, \href{http://dx.doi.org/10.1088/1126-6708/2009/03/139}
{J. High Energy Phys. {\bf 03} (2009), 139};
C.~Biggio, M.~Blennow, and E.~Fernandez-Martinez, \href{http://dx.doi.org/10.1088/1126-6708/2009/08/090}
{J. High Energy Phys. {\bf 08} (2009), 090}.

\bibitem{nsinuN}
J.~Barranco, O.G.~Miranda and T.I.~Rashba, \href{http://dx.doi.org/10.1088/1126-6708/2005/12/021}
{J. High Energy Phys. {\bf 0512} (2005), 021};
K.~Scholberg, \href{http://dx.doi.org/10.1103/PhysRevD.73.033005}
{Phys. Rev. D {\bf 73}, 033005 (2006)}.

\bibitem{gaitan}
R.~Gaitan \textit{et al.}, \href{http://dx.doi.org/10.1142/S0217751X13501248}
{Int. J. Mod. Phys. A {\bf28}, 1350124 (2013)}.

\bibitem{barranco2012}
J.~Barranco \textit{et al.}, \href{http://dx.doi.org/10.1142/S0217751X12501473}
{Int. J. Mod. Phys. A {\bf 27}, 1250147 (2012)}.

\bibitem{kjh2013}
K.J.~Healey, A.A.~Petrov, and D.~Zhuridov, \href{http://dx.doi.org/10.1103/PhysRevD.87.117301}
{Phys. Rev. D {\bf 87} (2013) 117301}.

\bibitem{dkptsk}
D.K.~Papoulias and T.S.~Kosmas, \href{http://dx.doi.org/10.1016/j.physletb.2015.06.039}
{Phys.Lett. B {\bf 747} (2015)}.

\bibitem{allen93}
R.C.~Allen \textit{et al.}, \href{http://dx.doi.org/10.1103/PhysRevD.47.11}
{Phys. Rev. D {\bf 47}, 11 (1993)}.

\bibitem{Georgi2007}
H.~Georgi, \href{http://dx.doi.org/10.1103/PhysRevLett.98.221601}
{Phys. Rev. Lett. {\bf 98}, 221601 (2007)};
H.~Georgi, \href{http://dx.doi.org/10.1016/j.physletb.2007.05.037}
{Phys. Lett. B {\bf 650}, 275 (2007)}.

\bibitem{Lenz2007}
A.~Lenz, \href{http://dx.doi.org/10.1103/PhysRevD.76.065006}
{Phys. Rev. D {\bf 76}, 065006 (2007)}.

\bibitem{Bolanos2007}
A.~Bolaos, \textit{et al.}, \href{http://dx.doi.org/10.1103/PhysRevD.87.016004}
{Phys. Rev. D {\bf 87}, 016004 (2013)}.

\bibitem{cheung2007}
K.~Cheung, W.Y.~Keung, and T.C.~Yuan, \href{http://dx.doi.org/10.1103/PhysRevLett.99.051803}
{Phys. Rev. Lett. {\bf 99}, 051803 (2007)};
K.~Cheung, W.Y.~Keung, and T.C.~Yuan, \href{http://dx.doi.org/10.1103/PhysRevD.76.055003}
{Phys. Rev. D {\bf 76}, 055003 (2007)}.

\bibitem{upchen07}
S.L.~Chen and X.G.~He, \href{http://dx.doi.org/10.1103/PhysRevD.76.091702}
{Phys. Rev. D {\bf 76}, 091702 (2007)}.

\bibitem{upbalantekin07}
A.B.~Balantekin and K.O.~Ozansoy, \href{http://dx.doi.org/10.1103/PhysRevD.76.095014}
{Phys. Rev. D {\bf 76}, 095014 (2007)}.

\bibitem{upbarranco09}
J.~Barranco \textit{et al.}, \href{http://dx.doi.org/10.1103/PhysRevD.79.073011}
{Phys. Rev. D {\bf 79}, 073011 (2009)}.

\bibitem{eagarces}
E.A.~Garc\'{e}s \textit{et al.}, \href{http://iopscience.iop.org/article/10.1088/1742-6596/378/1/012017}
{J. Phys. Conf. Ser. {\bf 378}, 012017 (2012)}.

\bibitem{hur2007}
T.I.~Hur, P.~Ko and X.H.~Wu, \href{http://dx.doi.org/10.1103/PhysRevD.76.096008}{Phys. Rev. D {\bf 76}, 096008 (2007)}.

\bibitem{atombinding}
S.A.~Fayans, L.A.~Mikaelyan and V.V.~Sinev, \href{http://dx.doi.org/10.1134/1.1398940}
{Phys. Atom. Nucl. {\bf 64}, 1475 (2001)};
M. B. Voloshin, \href{https://doi.org/10.1103/PhysRevLett.105.201801}
{Phys. Rev. Lett. {\bf 105}, 201801 (2010)},  {\bf 106}, 059901(E) (2011);
J.W.~Chen, H.C.~Chi, K.N.~Huang, C.P. Liu, H.T.~Shiao, L.~Singh , H.T.~Wong, C.L.~Wu and C.P.~Wu,
\href{http://dx.doi.org/10.1016/j.physletb.2014.02.036}{Phys. Lett. B {\bf 731}, 159 (2014)};
J.W.~Chen, H.C.~Chi, K.N.~Huang, H.B.~Li, C.P. Liu, L.~Singh , H.T.~Wong, C.L.~Wu and C.P.~Wu,
\href{https://doi.org/10.1103/PhysRevD.91.013005}{Phys. Rev. D {\bf 91}, 013005 (2015)};
J.W.~Chen, H.C.~Chi, S.T.~Lin, C.P. Liu, L.~Singh, H.T.~Wong, C.L.~Wu and C.P.~Wu,
\href{https://doi.org/10.1103/PhysRevD.93.093012}{Phys. Rev. D {\bf 93}, 093012 (2016)}.

\bibitem{grinstein08} B.~Grinstein, K.~Intriligator and I.~Z.~Rothstein, \href{http://dx.doi.org/10.1016/j.physletb.2008.03.020}
    {Phys. Lett. B {\bf 662}, 367 (2008)}.

\bibitem{upmontanino08}
D.~Montanino, M.~Picariello and J.~Pulido, \href{http://dx.doi.org/10.1103/PhysRevD.77.093011}
{Phys. Rev. D {\bf 77}, 093011 (2008)}.

\bibitem{Liao07}
Y. Liao and J.-Y. Liu, \href{http://dx.doi.org/10.1103/PhysRevLett.99.191804}
{Phys. Rev. Lett. {\bf 99}, 191804 (2007)}.

\bibitem{Gonzales08}
M.C. Gonzalez-Garcia, P.C. Holanda and R. Zukanovich Funchal,
\href{http://dx.doi.org/10.1088/1475-7516/2008/06/019}
{J. Cosmol. Astropart. Phys. 06 {\bf 19} (2008)}.

\bibitem{nuNcohsca}
H.T.~Wong \textit{et al.}, \href{http://dx.doi.org/10.1088/1742-6596/39/1/064}
{J. Phys. Conf. Ser. {\bf 39}, 266 (2006)}.

\end{thebibliography}
\end{document}